\documentclass[conference]{IEEEtran}
\IEEEoverridecommandlockouts
\usepackage{filecontents,lipsum}
\ifCLASSOPTIONcompsoc

   \usepackage{cite}
\else
   \usepackage{cite}
\fi
\ifCLASSINFOpdf
\usepackage[pdftex]{graphicx}
\else
\fi
\usepackage{amsmath}
\usepackage{amssymb}
\usepackage{array}
\graphicspath{ {Figures/} }
\usepackage{float}
\usepackage{multirow}
\usepackage{color}
\usepackage{lipsum}
\usepackage{mathtools}
\usepackage{cuted}
\usepackage{lipsum}
\usepackage{mathtools}
\usepackage{graphicx}
\usepackage[table]{xcolor}
\usepackage{nomencl}
\usepackage{amsmath}
\usepackage{tabularx}
\usepackage{booktabs}
\usepackage[table]{xcolor}
\usepackage{algorithm}
\usepackage{algpseudocode}
\usepackage{etoolbox}
\makeatletter
\newcommand*{\algrule}[1][\algorithmicindent]{%
  \makebox[#1][l]{%
    \hspace*{.2em}
    \vrule height .75\baselineskip depth .25\baselineskip
  }
}

\newcount\ALG@printindent@tempcnta
\def\ALG@printindent{%
    \ifnum \theALG@nested>0
    \ifx\ALG@text\ALG@x@notext
    \else
    \unskip
    \ALG@printindent@tempcnta=1
    \loop
    \algrule[\csname ALG@ind@\the\ALG@printindent@tempcnta\endcsname]%
    \advance \ALG@printindent@tempcnta 1
    \ifnum \ALG@printindent@tempcnta<\numexpr\theALG@nested+1\relax
    \repeat
    \fi
    \fi
}

\makenomenclature

\renewcommand{\figurename}{Fig.}
\usepackage{stfloats}
\usepackage{calc}
\newlength{\depthofsumsign}
\newlength{\totalheightofsumsign}
\newlength{\heightanddepthofargument}

\usepackage{nicematrix}
\usepackage{arydshln}
\usepackage{lipsum}

\usepackage{amsmath}

\usepackage{caption}
\usepackage{subfigure}
\usepackage{amsthm}

\renewcommand\IEEEkeywordsname{Keywords}
\renewcommand{\figurename}{Figure.}
\usepackage{textcomp}
\usepackage{bm}

\usepackage{xfrac}
 \usepackage{algorithm}
 \usepackage{algpseudocode}
 \usepackage{midfloat}
 \usepackage[none]{hyphenat}
\usepackage[justification=centering]{caption}	
\usepackage[labelsep=space]{caption}		
\usepackage[font=footnotesize]{caption}
\usepackage{float}

\begin{document}

\title{Adaptive Regulated Sparsity Promoting Approach for Data-Driven Modeling and Control of Grid-Connected Solar Photovoltaic Generation}

\author{Zhongtian Zhang,~Javad Khazaei,~\IEEEmembership{Senior Member,~IEEE},~and~Rick S. Blum~\IEEEmembership{Fellow,~IEEE.}

\thanks{This research was in part under support from the National Science Foundation under Grant NSF-EPCN 2221784. Zh. Zhang, J. Khazaei, and R. S. Blum are with the Electrical and Computer Engineering department at Lehigh University, PA, USA. (E-mails: $zhz819@lehigh.edu$, $jak921@lehigh.edu$, and $rblum@eecs.lehigh.edu$)}
}
\IEEEpubidadjcol
\IEEEpubid{\begin{minipage}{\textwidth}\ \\[25pt] \centering
  \color{blue}This work has been submitted to the IEEE for possible publication. Copyright may be transferred without notice, after which this version may no longer be accessible.
\end{minipage}}
\maketitle

\begin{abstract}  
This paper aims to introduce a new statistical learning technique based on sparsity promoting for data-driven modeling and control of solar photovoltaic (PV) systems. 
Compared with conventional sparse regression techniques that might introduce computational complexities  when the number of candidate functions increases, an innovative algorithm, named adaptive regulated sparse regression (ARSR) is proposed that adaptively regulates the hyperparameter weights of candidate functions to best represent the dynamics of PV systems. Utilizing this algorithm, open-loop and closed-loop models of single-stage and two-stage PV systems are obtained from measurements and are utilized for control design purposes. Moreover, it is demonstrated that the proposed data-driven approach can successfully be employed for fault analysis studies, which distinguishes its capabilities compared with other data-driven techniques. Finally, the proposed approach is validated through real-time simulations.

\end{abstract}

{\color{black}

}

\IEEEpeerreviewmaketitle


\section{Introduction}
 \IEEEPARstart{A}{vailability} of high-resolution measurements from field devices and existing challenges for accurately modeling distributed energy resources (DERs) has motivated data-driven modeling and control of smart grid assets.
 A wide range of system identification approaches have been introduced to extract dynamics from data. Some of these techniques include: dynamic mode decomposition (DMD) \cite{liu2020method,lu2020prediction}, Koopman operator \cite{al2021deep,mauroy2020koopman}, and sparse identification of nonlinear dynamics or sparse regression \cite{sindy,sindy_control}. These approaches have also been applied to power systems \cite{dmdp1, koopp, ddp1,li2022neural}. For instance, dynamic mode decomposition has been employed in \cite{dmdp1} for delay-tolerant microgrid control, while Koopman operator has been utilized in \cite{koopp} to identify generator dynamics for state estimation purposes.

The DMD is based on the assumption of linear dynamics and can not effectively handle nonlinear dynamics such as DERs. 
The Koopman operator is introduced as an infinite-dimensional linear operator that can sometimes 
 transform the dynamics of a nonlinear system into a linear representation. 
However, the transformation 
is only guaranteed in certain 
scenarios and under certain assumptions which do not generally apply \cite{al2021deep,mauroy2020koopman}. In contrast, sparse identification leverages sparse regression to identify the dominant dynamics of candidate functions and has demonstrated promising results in accurately modeling the unknown dynamics of nonlinear systems \cite{syndy,syndy1}. One of the key advantages of sparse regression is its implementation simplicity, reduced training time, interpretability, and superior performance compared to other model identification techniques. Numerous studies including our research demonstrated that sparse regression can be efficiently applied in the field of power and energy, including dynamics and impedance modeling of power converters in DC microgrids\cite{10372091}, data-driven nonlinear modeling and feedback linearization control of DERs\cite{ddsparse}, modularized sparse identification (M-SINDy) of microgrid transient dynamics\cite{msindy}, modeling of naval power systems\cite{ddnaval}, and power grid parameter estimation\cite{gpe}.

The existing literature underscores the substantial potential of sparse regression in identifying nonlinear dynamics within dynamical systems. However, photovoltaic (PV)-based DERs are instrumental in decarbonizing the power grid, yet a notable research gap exists in developing data-driven models and control strategies that cater to the diverse topologies and dynamics of solar photovoltaic systems, encompassing both single-stage and two-stage configurations. To the best of our knowledge, no existing study has explored data-driven modeling (both open-loop and closed-loop) and control of PV systems using sparse regression. 
Furthermore, a key  hyperparameter ($\lambda$ in (28)) 
 acts as a tuning knob for sparsifying the dynamics in conventional sparse regression methods is fixed and can cause model identification errors in high dimensional systems \cite{sindy,sindy_control,msindy}.The exploration of optimized selection of this key hyperparameter 
has not been addressed in existing research. 
 \begin{figure*}[htb]
    \centering
    \includegraphics[scale = 0.45]{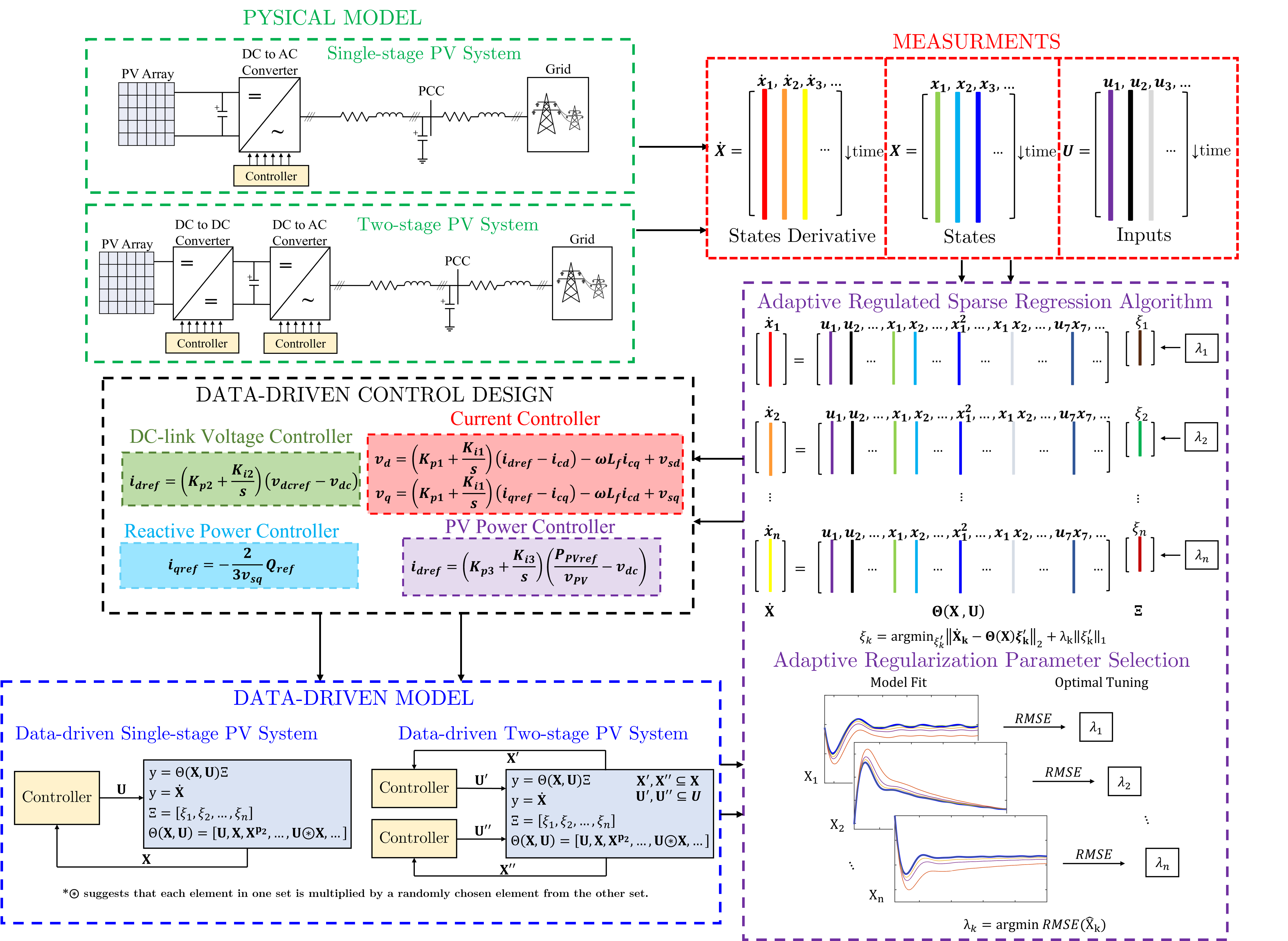}
    \caption{ Proposed adaptive regulated sparse regression for modeling identification and control of PV systems.}
    \label{dia} \vspace{-0.1in}
\end{figure*}

To address these limitations, this paper develops a novel data-driven modeling framework using adaptive regulated sparse regression (ARSR) for characterization of the dynamics exhibited by both single-stage and two-stage photovoltaic (PV) systems. 
The approach employs adaptive sparse regression to optimally identify the sparse dynamics from a library of candidate functions and construct accurate data-driven models using measurements. 
Next, an adaptive regularization approach is presented for optimal tuning of sparsity promoting hyperparameter, showcasing its capability to enhance the accuracy of data-driven modeling and enable the feasibility of closed-loop modeling for PV systems.
Finally, the paper introduces several practical applications of the proposed approach for 1) data-driven closed-loop control of PV systems, 2) detecting and analyzing faults, and 3) time-domain and real-time simulations demonstrate the efficacy of the ARSR for data-driven modeling and control of PV systems.

The rest of the paper is structured as follows: Section II formulates the problem. The ARSR method is introduced in Section III. Section IV covers the data-driven system modeling and control design. Section V comprises case studies, and Section VI concludes the paper.

\section{Problem Formulation}
This section presents the open-loop and closed-loop dynamic model of single-stage and two-stage PV systems to be utilized for the data-driven modeling approach in the next sections.  Fig. \ref{dia} shows the proposed data-driven modeling approach to solve this challenge. An  adaptive sparse regression algorithm is employed to identify the dynamics of PV systems using collected data.
Subsequently, the outcomes of the identification are harnessed to construct data-driven models for the PV systems. Following that, data-driven controllers are designed and implemented on the data-driven models.
\begin{figure}[htp]
\centering
\includegraphics[scale = 0.67]{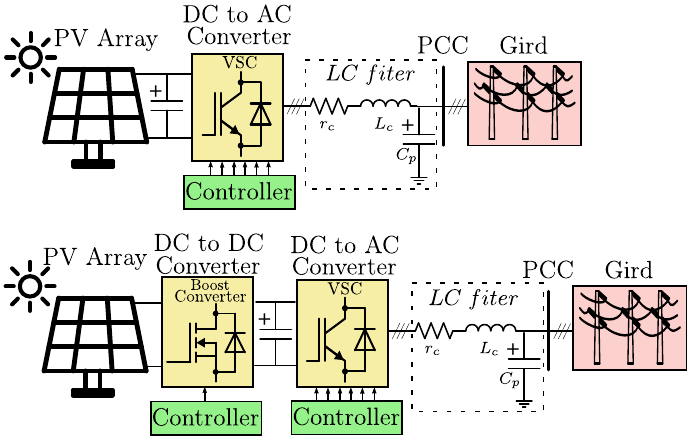}
\caption{Schematic of a single-stage and two-stage PV systems.  }
\label{fig1} \vspace{-0.1in}
\end{figure}
\subsection{Open-loop Dynamic Model and Control of PV systems}
\subsubsection{\textbf{Single-Stage PV}}
 Single-stage PV model consists of a PV array, a three-phase voltage source converter (VSC), and a low-pass filter connected to the main grid at a point of common coupling (PCC), as shown in Fig.\ref{fig1}. The DC-side of the VSC is connected to the PV array with a capacitor $C_{dc}$ and the AC-side is connected to an LC filter. The components of the filter are represented by $r_{c}$, $L_{c}$ and $C_{f}$, and the impedance of the grid is represented by $L_{g}$ and $r_{g}$, 
which refer to the inductance and resistance of the grid. 

\paragraph{AC-side Dynamics}
The AC-side dynamics of the system after converting to the dq-frame, can be described by 
\resizebox{.5\textwidth}{!}{
  \begin{minipage}{\linewidth}
\begin{align}
\begin{bmatrix}%
\dot{i}_{cd} \\
\dot{i}_{cq} \\
\dot{i}_{gd} \\
\dot{i}_{gq} \\
\dot{v}_{sd}\\
\dot{v}_{sq}
\end{bmatrix}
&=
\begin{bmatrix}
    -\dfrac{r_c}{L_c} & \omega_0  & 0        &  0 & -\dfrac{1}{L_c}  &0 \\
    -\omega_0 & -\dfrac{r_c}{L_c}  & 0        & 0 &  0  &-\dfrac{1}{L_c} \\
  \dfrac{1}{C_f} & 0  &  -\dfrac{1}{C_f}        & 0 & 0 &\omega_0 \\
    0 & \dfrac{1}{C_f}  & 0      & -\dfrac{1}{C_f} & -\omega_0    & 0 \\
     0 & 0  & -\dfrac{r_g}{L_g}  &   \omega_0     &  \dfrac{1}{L_g} & 0   \\
    0 & 0  & -\omega_0        &  -\dfrac{r_g}{L_g} & 0  &\dfrac{1}{L_g}  \\
\end{bmatrix} \begin{bmatrix}
i_{cd} \\
i_{cq} \\
i_{gd} \\
i_{gq} \\
v_{sd}\\
v_{sq}
\end{bmatrix}+ g(u) \notag
\end{align}
  \end{minipage}
}
\begin{align}
    g(u)&=\begin{bmatrix}
\dfrac{v_{cd}}{L_c} &
\dfrac{v_{cq}}{L_c}&
\dfrac{-v_{gd}}{L_g} &
\dfrac{-v_{gq}}{L_g} &
0&
0
\end{bmatrix}^T
\label{acdyn}
\end{align}
where the output voltage and current of the VSC are represented by $v_{cd}$, $v_{cq}$, $i_{cd}$ and $i_{cq}$ \cite{ddsparse}. Furthermore, the dq-frame voltages and currents of the grid are represented by $v_{gd}$, $v_{gq}$, $i_{gd}$ and $i_{gq}$. 
Also, the PCC voltage is represented by $v_{sd}$ and $v_{sq}$ in dq-frame, and $\omega_0$ is the nominal frequency of the grid.

\paragraph{DC-side Dynamics}
 Based on the power balance on the DC and AC sides of the VSC (steady state), the output power of the PV array ($P_{PV}$) should be equal to the VSC output power ($P_g$) \cite{single}. The voltage dynamic of the controller working on the DC-link capacitor can be described as
\begin{align}
    \frac{1}{2}C_{dc}v_{dc}\times v_{dc}s& = P_{PV}-P_g \label{vdc01} \\
   \dot{v}_{dc} &= \frac{i_{PV}}{C_{dc}}-\frac{3}{2}\frac{v_{gd}}{C_{dc}v_{dc}}i_{gd} \label{DCd} 
\end{align}
where $v_{gd}$ and $i_{gd}$ are the voltage and current of the grid {along the d-axis; $P_g=\frac{3}{2}v_{gd}i_{gd}$; $v_{dc}$ and $C_{dc}$ are the voltage and capacitance of the capacitor on the DC side of VSC; and $i_{PV}$ is the PV output current.}
\paragraph{State Space Model}
According to the AC-side and DC-side dynamics, the system can be represented in state space form as equations \eqref{acdyn} and \eqref{DCd}.

\begin{figure}[ht]
\centering
\includegraphics[scale = 1.0]{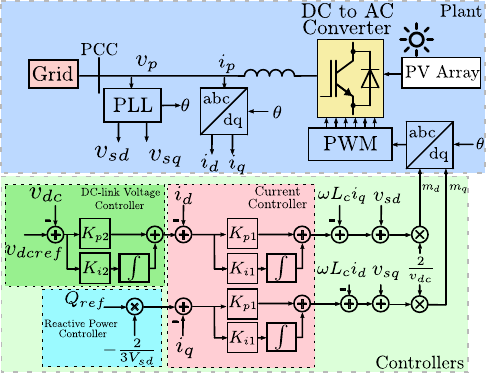}
\caption{The controllers for single-stage PV system.}
\label{ccon1}
\end{figure}
\paragraph{Current Controllers}
To regulate the converter current, a current controller is implemented (see Fig.\ref{ccon1}) with dynamics represented by  
\begin{align}
    v_{dref} = (K_{p1}+\frac{K_{i1}}{s})(i_{dref} - i_{cd})-\omega L_fi_{cq}+ v_{sd} \label{vrecd} \\
    v_{qref}= (K_{p1}+\frac{K_{i1}}{s})(i_{qref} - i_{cq})+\omega L_fi_{cd}+ v_{sq} \label{vrecq}
\end{align}
where $v_{dref}$ and $v_{qref}$ are the reference outputs in dq-frame while $i_{dref}$ and $i_{qref}$ are the reference currents of the controller in dq-frame \cite{two}. In addition, $K_{p1}$ and $K_{i1}$ are the proportional and integral parameters of the PI regulators in the current controller.

\paragraph{Active and Reactive Power Controllers}
In grid-tie PV systems, a phase-locked loop is implemented to synchronize  the converter frequency to the grid frequency by setting the q component of the converter voltage ($v_{cq}$) to 0 using a PI controller \cite{iravanibook}. Therefore, the dynamics of the active power $P$ and reactive power $Q$ delivered by the system to the grid is presented by 
\begin{align}
    i_{dref} = \frac{2}{3v_{sd}}P_{ref} ,  \quad
    i_{qref} = -\frac{2}{3v_{sq}}Q_{ref} \label{actP}
\end{align}
where the $P_{ref}$ and $Q_{ref}$ are the reference active and reactive powers \cite{two}.
\paragraph{DC-link Voltage Controllers}
In this case, since a DC-link controller is applied to the system, $i_{dref}$ is given by
\begin{align}
    i_{dref} = (K_{p2}+\frac{K_{i2}}{s})(v_{dcref} - v_{dc}) \label{VDCC}
\end{align} 
where $v_{dcref}$ is the reference of the DC-link voltage controller and $v_{dc}$ is the DC-link voltage. Parameters $K_{p2}$ and $K_{i2}$ are the proportional and integral parameters of the PI regulator of the DC-link controller \cite{two}.

In summary, a single-stage PV system is now represented by a nonlinear state space model of equations \eqref{acdyn} and \eqref{DCd}, incorporating two current controllers in 
equations \eqref{vrecd} and \eqref{vrecq}, a power controllers in equation \eqref{actP} , 
and a DC-link voltage controller in equation \eqref{VDCC}.

\subsubsection{\textbf{Two-stage PV}}
A two-stage PV system basically has a similar structure as a single-stage PV system with an additional DC to DC converter as shown in Fig.\ref{fig1}.
\paragraph{DC to DC Converter} The DC to DC converter amplifies the input voltage from the PV array by utilizing a maximum power point tracking (MPPT) controller, which delivers the target power reference ($P_{PVref}$) for the converter to track, as expressed in following equations \cite{two}.
\begin{align}
    \frac{di_{PV}}{dt} = \frac{1}{L_{b}}v_{PV}-\frac{(1-d_{ref})}{L_{b}}v_{dc} \label{ipv}\\
    \frac{dv_{dc}}{dt} = \frac{(1-d_{ref})}{C_{dc}}i_{PV}-\frac{1}{C_{dc}}i_{dc} \label{vdc2} 
\end{align}
where $v_{PV}$ and $i_{PV}$ are the output voltage and current of the PV array, $L_b$ is the inductor within the DC to DC converter, and $C_{dc}$ is the value of the capacitor on the DC side. The input current of the VSC, $i_{dc}$, can be represented by the power balance equation below\cite{two}.
\begin{align}
    v_{dc}i_{dc} &= \frac{3}{2}(v_{sd}i_{sd}+v_{sq}i_{sq})\label{balance}\\
    i_{dc} &= \frac{3}{2v_{dc}}(v_{sd}i_{sd}+v_{sq}i_{sq})\label{idc}
\end{align}
where $v_{sd}$, $v_{sq}$, $i_{sd}$ and $i_{sq}$ are the voltages and currents in dq-frame at the PCC.
In summary, the dynamic of a two-stage PV system is represented by linear dynamics in \eqref{acdyn} and nonlinear part shown below.
\begin{align}
    \dot{i}_{PV} &= \frac{1}{L_{b}}v_{PV}-\frac{(1-d_{ref})}{L_{b}}v_{dc} \label{ipv1}\\
    \dot{v}_{dc} &= \frac{(1-d_{ref})}{C_{dc}}i_{PV}-\frac{1}{C_{dc}}\frac{3}{2v_{dc}}(v_{sd}i_{sd}+v_{sq}i_{sq}) \label{vdc1} 
\end{align}

\begin{figure}[htb]
\centering
\includegraphics[scale = 0.95]{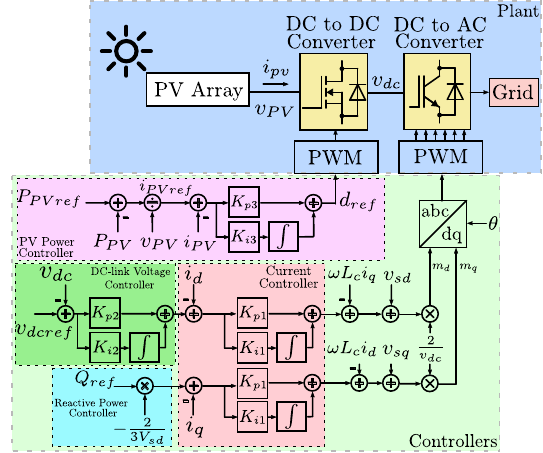}
\caption{Two-stage PV system controller.}
\label{pcon1}
\end{figure}

\paragraph{PV Power Controller}
As shown in Fig. \ref{pcon1}, besides the current controller, DC-link voltage controller and reactive power controller, a PV power controller is applied to the system. In equations \eqref{ipv} and \eqref{vdc2}, the new variable $d_{ref}$ is defined as the duty cycle reference of the PWM control, with dynamics represented by
\begin{align}
    d_{ref} = (K_{p3}+\frac{K_{i3}}{s})(i_{PVref} - i_{PV}) \label{ref}
\end{align}
where $i_{PV}$ is the output current of the PV array, 
while proportional parameter $K_{p3}$ and integral parameter $K_{i3}$ are the parameters of the PI regulators in the current controller. 
Given that $P_{PVref}=v_{PV}i_{PVref}$, equation \eqref{ref} can be rewritten as
\begin{align}
    d_{ref} = (K_{p3}+\frac{K_{i3}}{s})( \frac{P_{PVref}}{v_{PV}} - i_{PV}) \label{pp}
\end{align}
in which $P_{PVref}$ is the reference power value of the PV array, which is given by the MPPT controller, and $v_{PV}$ is the output voltage of the PV array. Since the dynamics of the MPPT controller is much slower than current and DC voltage controller, it is not considered for data-driven modeling \cite{two}. Instead, $P_{PVref}$ is assumed to be known. 

Now a two-stage PV system can be represented by a nonlinear state space model of equations \eqref{acdyn}, \eqref{ipv1} and \eqref{vdc1}, incorporating two current controllers in equations \eqref{vrecd} and \eqref{vrecq}, a DC-link voltage controller in equation \eqref{VDCC}, a reactive power ($Q$) controller, and a PV power controller in equation \eqref{pp}.
\subsection{Closed-loop Dynamic Model of PV systems}

 In an open-loop case, the PV plant model, excluding the controllers, will be replaced by a data driven model. In a closed-loop case, the combination of the PV plant and the controller with be replaced by a data-driven model. 
  \begin{figure}[htp!]
    \centering
    \includegraphics[scale = 0.9]{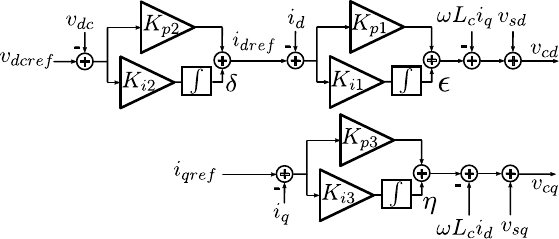}
    \caption{Controllers of the PV system.}
    \label{con}
\end{figure}
\paragraph{\textbf{Single-Stage System Controller Modeling}}
Fig. \ref{con} illustrates the structure of three PI controllers presented in equation \eqref{vrecd}, \eqref{vrecq}, and \eqref{VDCC}. 
For the modeling of controllers, the outputs of integrators are treated as the new states of the system, labeled as $\delta$, $\epsilon$ and $\eta$. The dynamics of these three states are expressed as
\begin{align}
   & \dot{\delta} = (v_{dcref}-v_{dc}) K_{i2}  \label{x8}\\
   & \dot{\epsilon} = ((v_{dcref}-v_{dc}) K_{p2} + \delta - i_d) K_{i1} \label{x9}\\  
   & \dot{\eta} = (i_{qef}-i_{q})K_{i3}\label{x10} 
\end{align}
The dynamic of the inverter's voltages $v_{cd}$ and $v_{cq}$ are then defined by the following equations
\begin{align}
   v_{cd} = &((v_{dcref}-v_{dc}) K_{p2} + \delta - i_d) K_{p1}  \notag\\
   &+\epsilon - L_c\omega_0 i_q + v_{sd} \label{vcdcl}\\  
    v_{cq} = &((i_{qef}-i_{q}) K_{p3} + \eta + L_c \omega_0  i_d+ v_{sq}\label{vcqcl}
\end{align}
A closed-loop single-stage PV system can be represented by a nonlinear state space model of equations \eqref{acdyn}, \eqref{DCd}, \eqref{x8}, \eqref{x9} and \eqref{x10}. 
\section{Sparse Identification of PV Systems}
 Consider a dynamic system, i.e. a open-loop single-stage PV system represented by equations \eqref{acdyn} and \eqref{DCd}, a open-loop two-stage PV system represented by equations \eqref{acdyn}, \eqref{ipv1} and \eqref{vdc1} or a closed-loop single-stage PV system represented by \eqref{acdyn}, \eqref{DCd}, \eqref{x8}, \eqref{x9} and \eqref{x10}, following the general equation:
\begin{align}
    \dot{\mathbf{x}}(t)=\mathbf{f}(\mathbf{x}(t),\mathbf{u}(t)). \label{sys}
\end{align}
Here the state of a system at time $t$ is represented by the vector $\mathbf{x}(t)\in \mathbb{R}^n$  and the inputs to the system at time $t$ are represented by $\mathbf{u}(t)\in \mathbb{R}^m$, where $n$ and $m$ denote the dimensions of the state and inputs. In this paper, the states and inputs of a open-loop single-stage PV system are denoted as follows: $X_1 = [i_{cd},i_{cq},i_{gd},i_{gq},v_{sd},v_{sq},v_{dc}]$ and $U_1 = [v_{cd},v_{cq},v_{gd},v_{gq},\omega_0,i_{PV}]$. For a open-loop two-stage PV system, the states and inputs are represented as: $X_2=[i_{cd},i_{cq},i_{gd},i_{gq},v_{sd},v_{sq},v_{dc},i_{PV}]$ and $U_2=[v_{cd},v_{cq},v_{gd},v_{gq},\omega_0,v_{PV},d_{ref}]$. In the case of a closed-loop single-stage PV system, the states and inputs are specified as: $X_3=[i_{cd},i_{cq},v_{gd},v_{gq},v_{sd},v_{sq},v_{dc},\delta,\epsilon,\eta]$ and $U_3 = [v_{dcref},i_{qref},v_{gd},v_{gq},i_{PV}$].   The goal in this section is to identify a data-driven version of the function $\mathbf{f}(\mathbf{x}(t),\mathbf{u}(t))$ represented as $\mathbf{\dot{X}}=\Theta(X,U)\Xi$ for the PV systems from available data. This process is explained in the following.
\subsection{Data Collection}
According to \eqref{sys}, the time history of states $\mathbf{x}(t)$ and inputs $\mathbf{u}(t)$ shall be collected\cite{syndy}, and the derivative $\dot{\mathbf{x}}(t)$ can either be measured or numerically approximated from $\mathbf{x}(t)$. The data then are sampled 
and organized as three matrices
\begin{align}
    \mathbf{X}&=
    \begin{bmatrix}
    \vline    & \vline &  & \vline\\
    {x_1}(t_k)    & {x_2}(t_k) & \cdots & {x_n}(t_k)\\
    \vline    & \vline &  & \vline\\
    \end{bmatrix}time\downarrow \label{x}\\
    \mathbf{U}&=
    \begin{bmatrix}
    \vline    & \vline &  & \vline\\
    {u_1}(t_k)    & {u_2}(t_k) & \cdots & {u_m}(t_k)\\
    \vline    & \vline &  & \vline\\
    \end{bmatrix}time\downarrow\label{u}\\
    \dot{\mathbf{X}}&=
    \begin{bmatrix}
    \vline    & \vline &  & \vline\\
    \dot{x}_{1}(t_k)    & \dot{x}_{2}(t_k) & \cdots & \dot{x}_{n}(t_k)\\
    \vline    & \vline &   & \vline\\
    \end{bmatrix}time\downarrow \label{dx}
\end{align}
\subsection{Library Construction}
The next step is to construct a library $\Theta(\mathbf{X}, \mathbf{U})$ containing all the candidate functions of the variables in the columns of $\mathbf{X}$ and $\mathbf{U}$, as shown in equation \eqref{theta} and \eqref{xp2},
\begin{figure*}[htb]
\begin{align}
    \Theta(\mathbf{X,U}) =
    \begin{bmatrix}
     \vline &\vline & \vline & \vline&\vline & \; & \vline&\vline & \;\\
     1 &U  & \mathbf{X} & \mathbf{(X,U)}^{P_2}&\mathbf{(X,U)}^{P_3}  & \cdots & sin(\mathbf{(X,U)})& cos(\mathbf{(X,U)})& \cdots\\
     \vline &\vline & \vline & \vline&\vline & \; & \vline&\vline & \;
    \end{bmatrix}time\downarrow \label{theta}
\end{align} \vspace{-0.1in}\\
\begin{align}
    \mathbf{(X,U)}^{P_2}=
    \begin{bmatrix}
    \vline&\vline &  &\vline  &  & \vline&\vline&\vline&  &\vline\\
    \mathbf{u_1^2}(t_k)&\mathbf{u_1}(t_k)\mathbf{u_2}(t_k) & \cdots&\mathbf{u_1}(t_k)\mathbf{x_1}(t_k)& \cdots& \mathbf{u_m^2}(t_k)&\mathbf{x_1^2}(t_k)&\mathbf{x_1}(t_k)\mathbf{x_2}(t_k)& \cdots&\mathbf{x_n^2}(t_k)s\\
    \vline&\vline & &\vline& & \vline&\vline&\vline&  &\vline
    \end{bmatrix}\label{xp2}
\end{align}
\end{figure*}
where $\mathbf{X}^{P_2}$ and $\mathbf{X}^{P_3}$ represent high degree polynomials. For example, $\mathbf{X}^{P_2}$, with second order terms, is shown in equation \eqref{xp2}.
During the regression process, the construction of the library can significantly influence the identification results, the identification process becomes more accurate when there is a greater understanding of the system and reasonable terms are carefully selected in the library.
 \vspace{-0.2cm}
\subsection{Sparse Regression}
Each column of $\Theta(\mathbf{X}, \mathbf{U})$ represents a potential candidate function for the right-hand side of equation \eqref{sys}. The matrix $\Theta(\mathbf{X}, \mathbf{U})$ allows considerable flexibility in selecting the specific nonlinearities for inclusion in the candidate functions. Since it is believed that only a small number of these nonlinearities are active in each row of $\mathbf{f}(\mathbf{x}(t),\mathbf{u}(t))$, a sparse regression problem can be formulated to identify the sparse coefficient vectors $\Xi =[\xi_1,\xi_2,\hdots,\xi_k,\hdots,\xi_n]$ indicating which nonlinearities are active. The process can be articulated as an optimization problem
\begin{align}
    \xi_k = \operatorname*{argmin}_{\xi_{k}'}  \bigg\lVert\mathbf{\dot{X}_k-\Theta(X,U)\xi_{k}'}\bigg\rVert_{2}+ \lambda \bigg\lVert \xi_{k}'\bigg\rVert_0
    \label{sindy11}
    \end{align}
here $\left|\left|\cdot\right|\right|_{2}$ denotes the $L2$-norm, measuring the overall magnitude of the vector, $\left|\left|\mathbf{\dot{X}_k-\Theta(X,U)\xi_{k}'}\right|\right|_{2}$ represents the optimization problem aiming to find the coefficient vector $\xi_{k}$ that minimizes the disparity between the measured data $\dot{X}_k$ and the data-driven model $\Theta(X,U)\xi_{k}'$. In addition, $\left|\left|\cdot\right|\right|_{0}$ denotes the $L0$-norm, the number of nonzero elements in the coefficient vector \cite{sindy}. The parameter $\lambda$ serves as the trade-off between fitting the data well and promoting sparsity in the solution. It penalizes large values 
and encourages many coefficients to be exactly zero. Selecting an appropriate value for $\lambda$ is crucial in the sparse regression approach, which is thoroughly discussed in section V.A. 

\subsection{Adaptive Regulated Sparse Regression}
In order to better evaluate the performance of the method, we first introduce our optimization criteria, Root Mean Square Error (RMSE), which measures the error between the true and predicted values of the $k$-th state, which is denoted as:
\begin{align}
    RMSE\{Data(\Xi),Phy \}_k =\sqrt{\frac{\sum_{i=1}^P (y_{kpi}-y_{kdi})^2}{P}} \label{RMSE}
\end{align}
where $P$ represents the number of data points, $y_{kpi}$ denotes the measurement of $k$-th state of the physical model at data point $i$, and $y_{kdi}$ represents the prediction  of $k$-th state of the data-driven model for data point $i$. The RMSE values serve as a quantitative measure of the dissimilarity between the physical model and the data-driven model outputs, providing insight into the accuracy and performance of the data-driven model.  

Conventional sparse regression approaches employ one single $\lambda$ value for the entire model, which poses challenges in managing complex models such as the closed-loop model of PV systems. 
This challenge arises because the coefficients of distinct terms in the differential equations for each state may exhibit notable variations. For instance, in equation \eqref{x8} for $\delta$, the coefficient $K_{i2}$ is 0.01, while in equation \eqref{x9} describing the dynamic of $\epsilon$, the coefficient $K_{i1}$ is 500. Identifying such coefficients proves to be difficult using the same threshold as in the previous section. To address this issue, a new algorithm has been developed, which is described as an optimization problem in the following
\begin{align}
   &\xi_k = \operatorname*{argmin}_{\xi_{k}'}  \bigg\lVert\mathbf{\dot{X}_k-\Theta(X)\xi_{k}'}\bigg\rVert_{2}+ \lambda_k \bigg\lVert \xi_{k}'\bigg\rVert_0
    \label{sindy12}\\
    &\lambda_k = \operatorname*{argmin}_{\lambda_k}  RMSE\{Data(\Xi),Phy\}_k
    \label{sindy22}
    \end{align}
where $\xi_k$ is the $k$-th column of coefficient matrix $\Xi$, $\dot{X}_k$ is the derivative of $k$-th state and $\lambda_k$ is the $k$-th regularization parameter. 
The key enhancement lies in the algorithm's capability to adapt the regularization parameter $\lambda$ dynamically while identifying distinct states of the system. As shown in equation \eqref{sindy12}, parameter $\lambda$ is now represented as a vector $\mathbf{\Lambda} =[\lambda_1,\lambda_2,\hdots,\lambda_k,\hdots,\lambda_n]$, where the number of elements corresponds to the number of states and the parameter for $k$-th state, $\lambda_k$, is determined by equation \eqref{sindy22}. The sparse regression method outlined in equation \eqref{sindy12} is denoted as $SparseRegression\{\mathbf{\dot{X}},\Theta(\mathbf{X,U}),\mathbf{\Lambda}\}$.

Algorithm \ref{alg:lambdaSINDy} presents the adaptive regulated sparse regression, defined as \textbf{AdaptiveSINDy}$\mathbf{\{\dot{X},\Theta(X,U)\}}$, which can be summarized in three main steps: 1) Initialization, 2) Data-driven modeling, and 
 3) RMSE calculation and Optimization loop. It is crucial to emphasize that to compute the RMSE, a data-driven model should be constructed using training data (discussed in Section IV). Next we calculate the RMSE between the measured states and data-driven model predictions and 
 compare them with the test data. Moreover, since the relationship between regulation parameters and state errors is non-linear, the algorithm updates the parameters using a predefined step unit until reaching the minimum state error.
The performance of the ARSR is discussed in section V.A and V.B.

\begin{algorithm}[htp]
	\caption{Adaptive Regulated Sparse Regression} \small
	\label{alg:lambdaSINDy}
 \hspace*{\algorithmicindent} \textbf{Inputs:} Measured                derivatives$\mathbf{\dot{X}}$, library $\Theta(\mathbf{X,U})$, Optimized\\
 \hspace*{\algorithmicindent} 
 \textbf{\;\;\;\;\;\;\;\;\;\;\;\;\;}states array $StateX$\\
      \hspace*{\algorithmicindent} \textbf{Output:}  Sparse regression model coefficients $\Xi$, Optimal\\
      \hspace*{\algorithmicindent} \textbf{\;\;\;\;\;\;\;\;\;\;\;\;\;\;}regularization parameter vector $\Lambda$.     
        \begin{algorithmic}[]
         \raggedright 
        
        \State \textit{Initialization} :
        \State $\Lambda = [\lambda_1,\lambda_2,...,\lambda_n];$ \Comment{Init $\lambda s$ array}
        \State $\lambda_{max}$, $Steps;$ \Comment{set the upper limit and unit steps}
         
         \end{algorithmic} 
          \begin{algorithmic}[1]
         \raggedright 
       \State $\textbf{AdaptiveSINDy}\{\mathbf{\dot{X}},\Theta(\mathbf{X,U})\}$

	\State $\hat{\Xi} = SparseRegression\{\mathbf{\dot{X}},\Theta(\mathbf{X,U}),\mathbf{\Lambda}\};$ \Comment{find $\hat{\Xi}$ with initial $\lambda$s}
	\State $err =RMSE\{ Data(\hat{\Xi}), Phy\};$ \Comment{find the RMSEs for each state between physical model and data-driven model with current coefficient matrix $\hat{\Xi}$}
	\State $Ind =max(err) $ \Comment{find the state with the largest RMSE}
        \While{$Ind \leq n$} \Comment{$n$ is number of states}
            \State $\hat{\Lambda} = \Lambda$; \Comment{update temporary lambda array}
           
            \For{$\hat{\Lambda}(Ind)=\Lambda(Ind) :Steps(Ind):  \lambda_{max}(Ind) $}
                        
                \State $\hat{\Xi} = SparseRegression\{\mathbf{X},\mathbf{\dot{X}},\Theta(\mathbf{X}),\hat{\Lambda} \};$
                \State $err =RMSE\{ Data(\hat{\Xi}), Phy\};$
	           \If {$err(Ind)$ is reduced}
                        \State $\Lambda(Ind)= \hat{\Lambda}(Ind)$
                        \EndIf
            \EndFor
            \State$ StateX \leftarrow Ind$ \Comment{update $StateX$ with optimized states}
            \State $Ind = max(err)$
        
        \While{$Ind\in StatesX$}
                \State $Ind = nextmax(err)$ \Comment{find the state with next largest RMSE}
                    \If {$length(StateX)=n$} 
                        \State $\Xi = \hat{\Xi};$ \Comment{update coefficient matrix}
                        \State \textbf{return}  \Comment{The algorithm ends with all states are optimized}
                        \EndIf
            \EndWhile
        \EndWhile    
	\end{algorithmic} 
\end{algorithm}
\section{Data-Driven Modeling and Control}
Using the outcomes of the proposed ARSR, it becomes feasible to construct data-driven controllers for both open-loop and closed-loop PV systems. 
\begin{figure*}[ht]
\centering
\includegraphics[width = 0.9\textwidth]{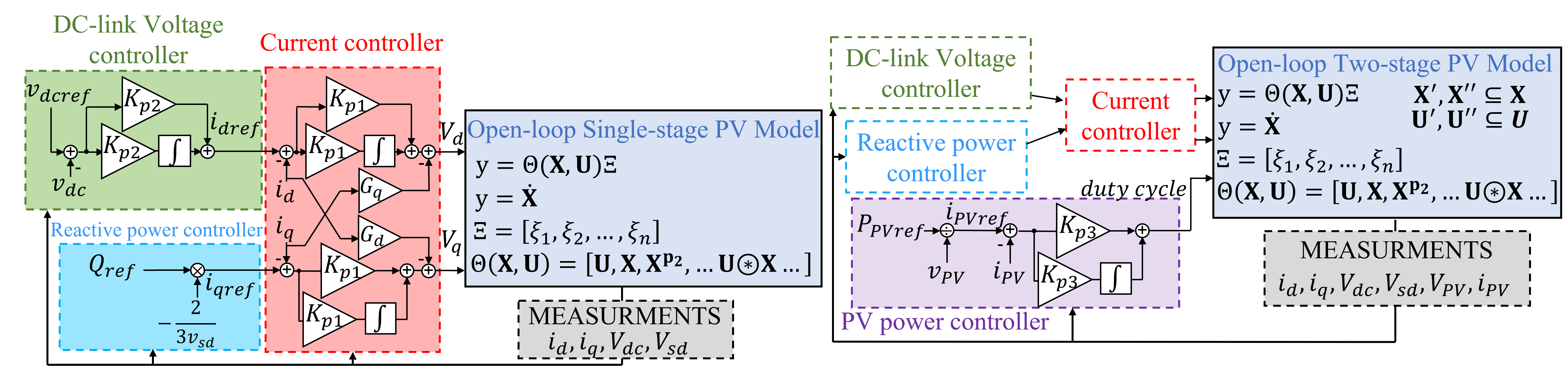}
\caption{Data-driven control design of single-stage and two-stage PV systems.}
\label{data2}
\end{figure*}
\subsection{Open-loop Data-driven Modeling with Controllers} As shown in Fig.\ref{data2}, the control of the physical model of both PV system consists of a DC-link voltage controller, a reactive power controller and a current controller in dq-frame, while the two-stage PV system has an additional controller for regulating the PV power via DC to DC converter. The dynamics of the data-driven DC-link voltage controller and the reactive power controller should mirror those of the physical model. The outputs of the data-driven controllers will decide the converter voltages in dq frame.
\subsection{Data-driven Control Design}

\begin{figure}[ht]
\centering
\includegraphics[scale = 1.0]{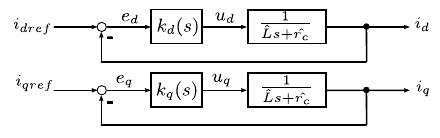}
\caption{dq-frame Closed-loop Current Controller.  }
\label{ccon2}
\end{figure}

The design of the parameters in the PI controllers for these current controllers can now be accomplished based on the identified data-driven model using ARSR. As shown in Fig. \ref{ccon2}, a simplified block diagram of current control loop is utilized for designing the bandwidth of controllers in the data-driven model\cite{iravanibook}.\\
The data-driven plant dynamics are denoted by $\hat{L}$ and $\hat{r_c}$, which can be obtained through the sparse regression algorithm, as specified in equation \eqref{acdyn}. Taking the d-axis controller as an example, the data-driven current controller $k_d(s)$ can be expressed as
\begin{align}
    k_{d} = \frac{k_ps+k_i}{s} \label{pid}
\end{align}
where $k_p$ and $k_i$ denote the data-driven proportional and integral gains of the current controller, respectively. Then the loop gain is expressed as
\begin{align}
    l(s) = \biggr{(}\frac{k_p}{Ls}\biggr{)}\frac{s+k_i/k_p}{s+r_c/L} \label{loopgain}
\end{align}
Due to the pole of the loop gain at $s=-r_c/L$, an ideal loop gain takes the form $l(s) = k_p/Ls$, where the pole is cancelled by the compensator zero at $s=-k_i/k_p$. Consequently, the closed-loop transfer function becomes
\begin{align}
    G(s) = \biggr{(}\frac{l(s)}{1+l(s)}\biggr{)}=\frac{k_p/L}{s+k_p/L}=\frac{1}{\tau_is+1} \label{clt}
\end{align}
where $\tau_i$ is the time constant of the current controller represented by $\tau_i=L/k_p$. Thus the data-driven proportional and integral gains $k_p$ and $k_i$ are obtained.
\begin{align}
    k_p = \frac{L}{\tau_i},\; k_i = \frac{r_c}{\tau_i} \label{kpi}
\end{align}
Similar concept can be used to obtain the data-driven control design for other controllers such as DC voltage and power controller. Utilizing the constructed open-loop models and controllers, data-driven single-stage and two-stage PV systems can be developed based on the outcomes of the ARSR algorithm. 

 \section{Case Studies}
 To evaluate the effectiveness of the proposed ARSR approach for modeling and control of PV systems, various case studies are conducted using time-domain  and real-time simulations in MATLAB and Opal-rt. 
 \subsection{Impact of hyperparameter $\lambda$ for PV systems}
 This section demonstrates the impact of various $\lambda$ values on the identification of open-loop models for both single-stage and two-stage PV systems.
 \subsubsection{\textbf{Impact of various $\lambda$ in single-stage PV models}}
\begin{figure}[htb!]
    \centering
   
    \includegraphics[width=0.35\textwidth]{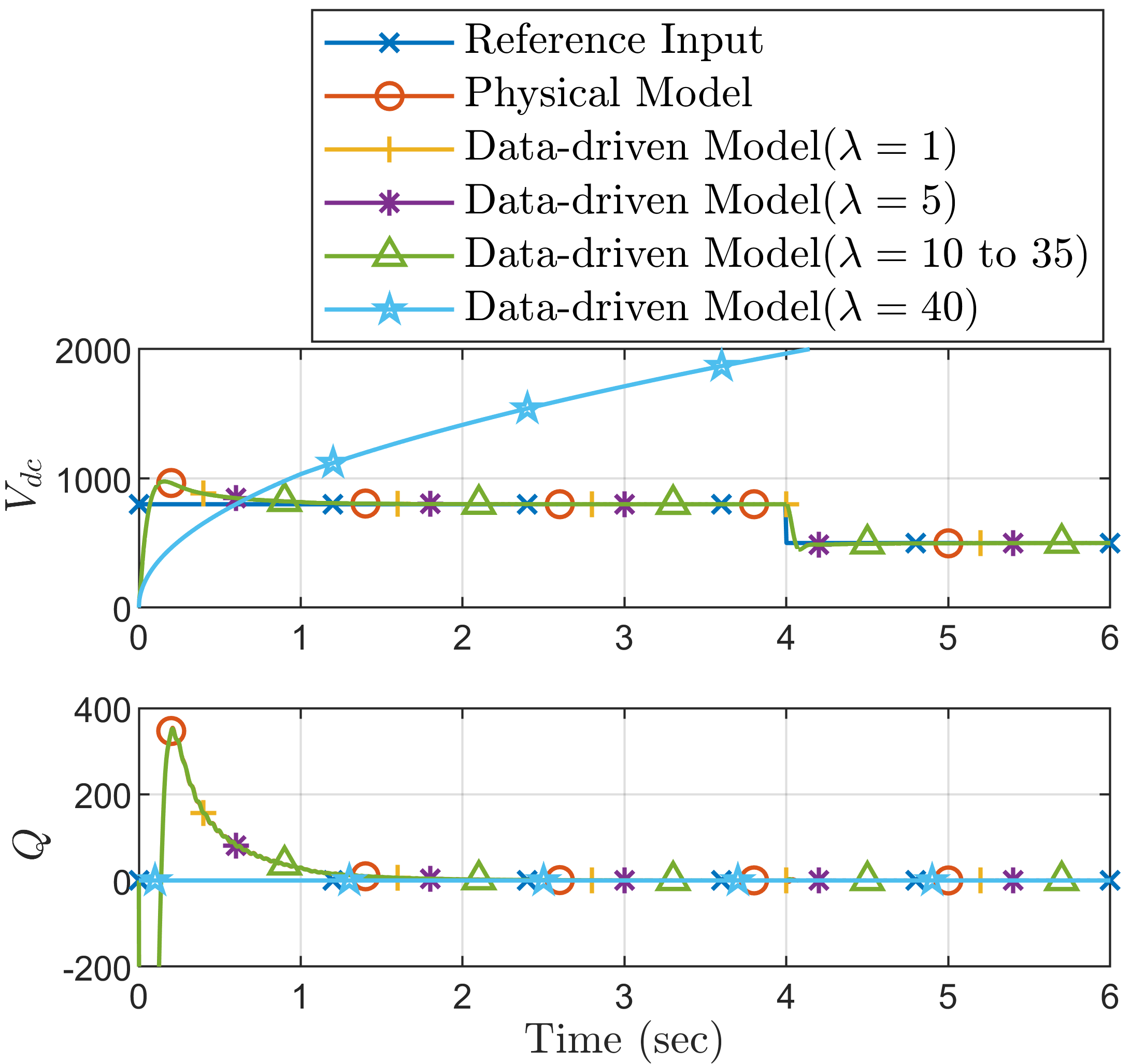}
    
    \caption{Output signals comparisons of single-stage PV system.}
    \label{xxx}
\end{figure}

 Fig. \ref{xxx} and the left part of Table \ref{rmse1} illustrates the tracking performance of the single-stage PV system, which comprise two reference inputs: the DC-link voltage $v_{dc}$  and reactive power $Q$. As depicted in the figure, the data-driven models closely approximate the physical model in the majority of cases when $\lambda$ ranges from 1 to 35. Notably, in Table \ref{rmse1}, when $\lambda$ ranges from 5 to 35 the model achieves the best performance.
\subsubsection{\textbf{Impact of Various $\lambda$ in Two-stage PV Models}}
\begin{figure}[htb!]
    \centering
    \includegraphics[width=0.38\textwidth]{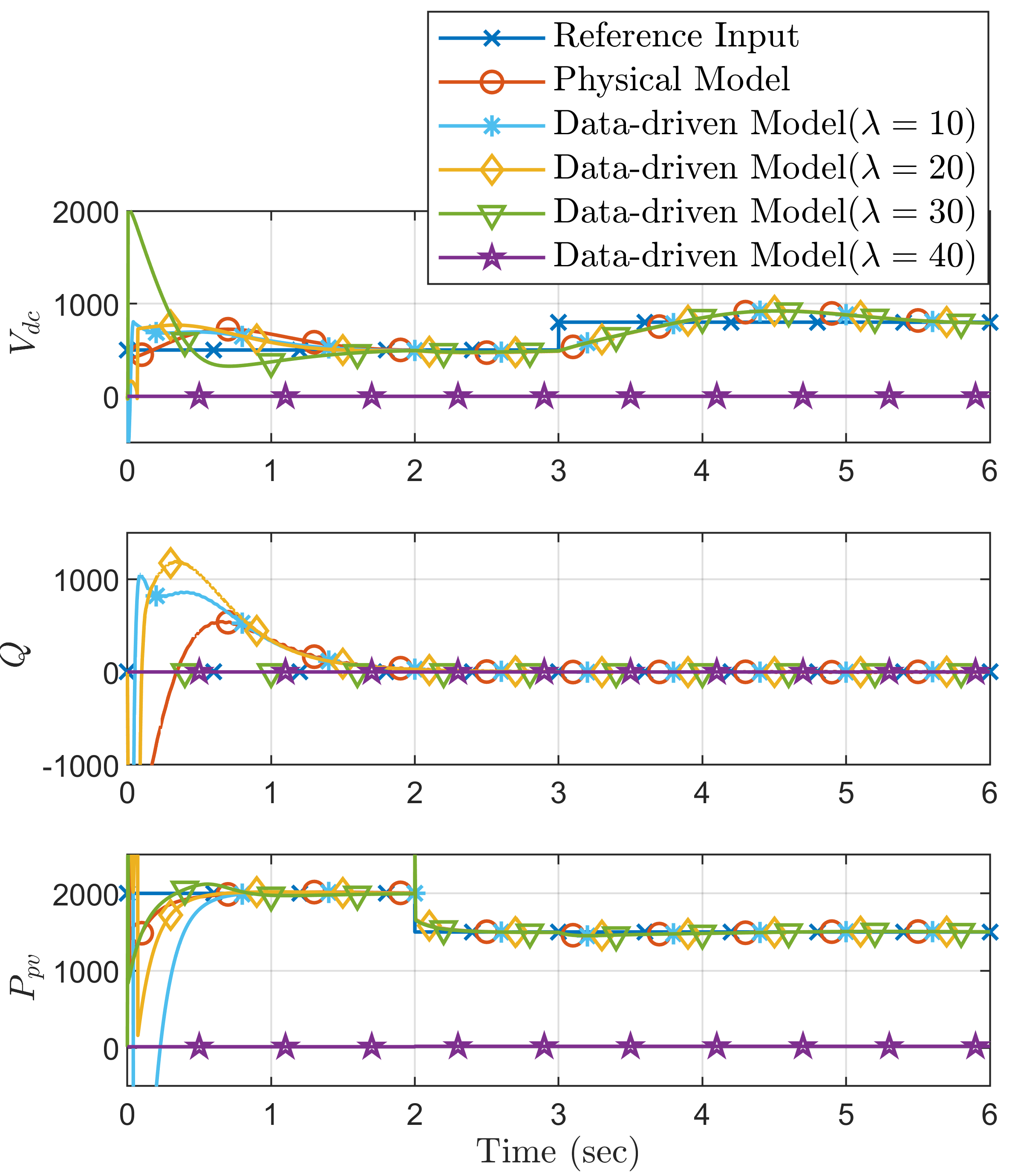}
    \caption{Output signals comparisons of two-stage PV system.}
    \label{x1}
\end{figure}

Fig. \ref{x1} and right part of the Table \ref{rmse1} illustrate the tracking performance of the two-stage PV model, which comprises three reference inputs: the DC-link voltage $v_{dc}$, reactive power $Q$, and  PV array power $P_{PV}$. When $\lambda$ ranges from 1 to 30, the data-driven models closely approximate the physical model. It is observed that the variables $v_{dc}$ exhibits optimal performance when $\lambda$ is set to 30. On the other hand, the variable $P_{PV}$ and $Q$ demonstrate its best performance when $\lambda$ falls to 5 and 1, respectively.

\subsubsection{\textbf{Analysis of Adaptive Method}}
To further improve the results, the adaptive method outlined in section III.D is applied to the open-loop modeling process for both single-stage and two-stage PV systems. The resulting regulation parameters are $\mathbf{\Lambda_1} =[5,5,5,5,30,30,20]$ and $\mathbf{\Lambda_2} =[1,5,5,5,30,30,1,35]$, respectively. Examining the tracking performance in the last row of Table \ref{rmse1}, it is evident that the single-stage models do not exhibit improvement since the conventional method already achieves optimal performance. However, the two-stage models demonstrate a significant enhancement, indicating that the adaptive method yields better results in modeling complex systems.
\begin{table}[htbp]
\caption{References tracking performance(RMSEs) of single-stage and two-stage PV systems} \vspace{-0.1in}
\begin{center}
\begin{tabular}{cccccc}
\hline
\hline
\multirow{3}{*}{\textbf{$\lambda$}}& \multicolumn{2}{c}{\textbf{\textit{Single-stage}}} & \multicolumn{3}{c}{\textbf{\textit{Two-stage}}}\\ \cmidrule(lr){2-3} \cmidrule(lr){4-6}

 & \textit{ $v_{dc}$ error}& \textit{$Q$ error}& \textit{ $v_{dc}$ error}& \textbf{\textit{$Q$ error}}& \textbf{\textit{$P_{PV}$ error}}\\
\hline\vspace{-0.2cm}\\
            1         & 11.3752	&28.9546 & 39.9779	&30.7711	&212.33\\
                5        &1.2882	&3.2921   & 39.7971	&27.3171	&296.048\\
                10       &1.2882	&3.2921 & 115.5112	&59.6551	&1930.4\\
                15        &1.2882	&3.2921  & 98.9295	&51.8175	&1913.9\\
                20        &1.2882	&3.2921   & 98.9295	&51.8175	&1913.9  \\
                25        &1.2882	&3.2921   & 98.9295	&51.8175	&1913.9\\
                30       &1.2882	&3.2921	&36.9085
	&339.3858	&403.3931  \\
                35       &1.2882	&3.2921	&36.9085
	&339.3858	&403.3931    \\
                40        &644.5295	&358.758&696.4531	&339.3858	&1651.7\\
Adpt.& 1.2882&3.2921 & 36.9085& 26.6831&6.1833\\
\hline
\hline
\end{tabular}
\label{rmse1}
\end{center}
\end{table}

In conclusion, the results obtained from the identification process in this section demonstrate that varying the values of $\lambda$ can significantly impact the performance of the sparse regression approach. 
 \begin{table*}[htbp]
\caption{Regulation parameters and state errors closed-loop single-stage PV system }\vspace{-0.1in}
\begin{center}
\begin{tabular}{ccccccccccc}
\hline
\hline
& \textbf{$x_1$}& \textbf{$x_2$}& \textbf{$x_3$}& \textbf{$x_4$}& \textbf{$x_5$}&\textbf{$x_6$}&\textbf{$x_7$}&\textbf{$x_8$}&\textbf{$x_9$}&\textbf{$x_{10}$}   \\ \hline 
                $\lambda$ & 1	&1	&2	&1	&7	&19	&1&	0.01&	1&	1\\ \hline\vspace{-0.2cm}\\
                State Error      &0.0409	&0.0006	&0.0244	&0.1548	&1.5791	&0.2592	&0.8664	&0.0009&0.0181	&0.0008\\ 
\hline
\hline
\end{tabular}
\label{tablermse}
\end{center}
\end{table*}
 \begin{figure}[htb!]
    \centering
    \includegraphics[width=0.35\textwidth]{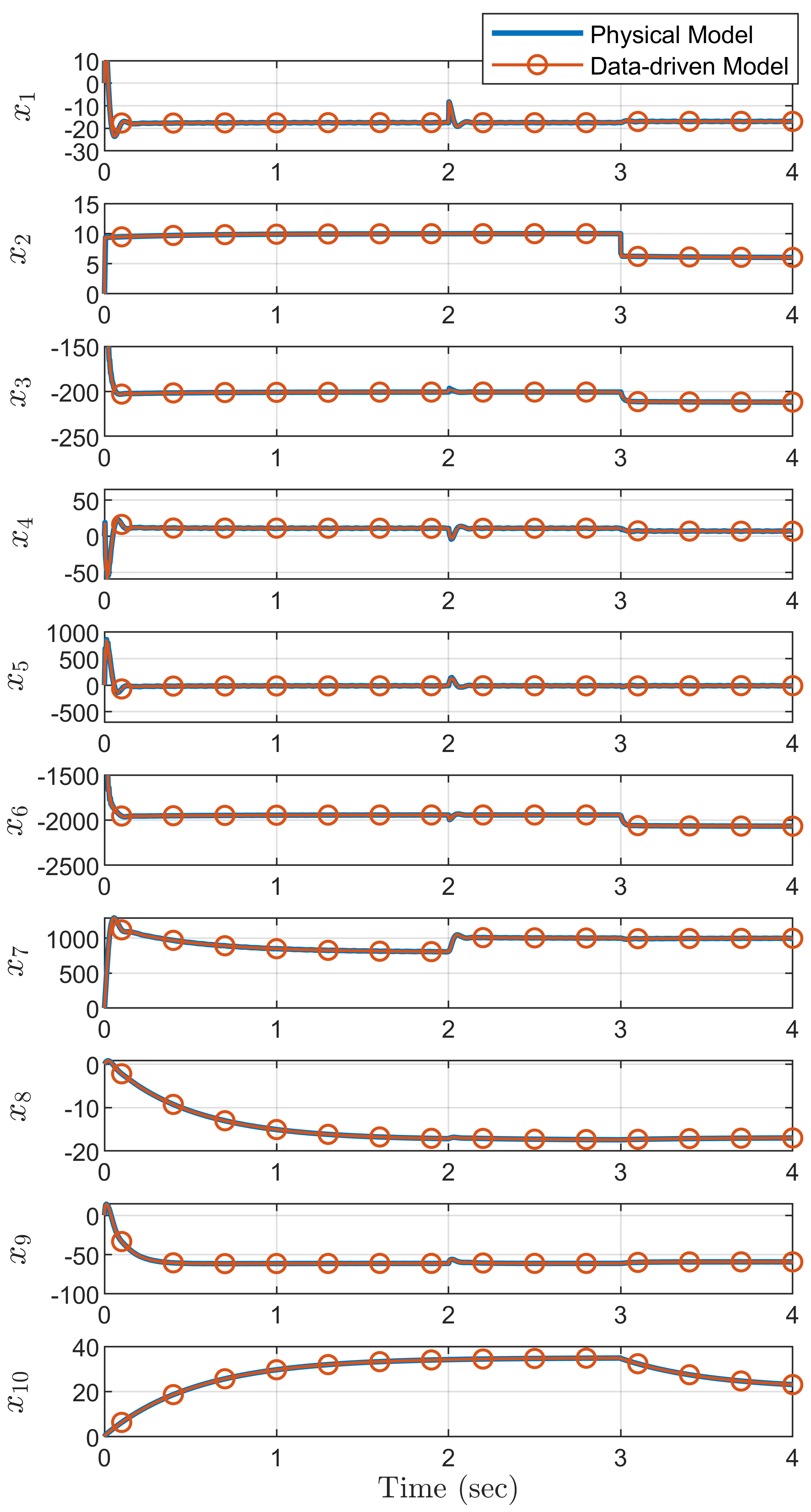}
    \caption{States comparisons of physical model and closed-loop data-driven model.}
    \label{cl1}
\end{figure}
\subsection{Closed-loop Modeling for Single-stage PV Systems}
This case study primarily addresses the closed-loop model identification of a single-stage PV system using the proposed ARSR algorithm. After employing measured data from the single-stage PV system and utilizing the obtained sparse matrix of coefficients $\Xi$ via ARSR, the data-driven model is formulated. To validate its effectiveness, a comparison is conducted with the physical model. The outcomes of regulation parameters $\lambda$ and state errors (RMSEs) are presented in Table \ref{tablermse} and Fig. \ref{cl1}. The rationale for selecting states to assess the system is rooted in the fact that, unlike the open-loop model with controllers, which involves multiple reference inputs for the controllers to regulate, in the closed-loop model, it is more appropriate to compare all the states of the system. The results illustrate that the data-driven model exhibits high accuracy when compared to the original physical model, a conclusion further supported by the small state errors detailed in Table \ref{tablermse}.
The outcomes of the closed-loop modeling in this section demonstrate the effectiveness of the ARSR algorithm in close-loop sparse regression modeling for a single-stage PV system.

\begin{figure}[htb!]
    \centering
    \includegraphics[width=0.4\textwidth]{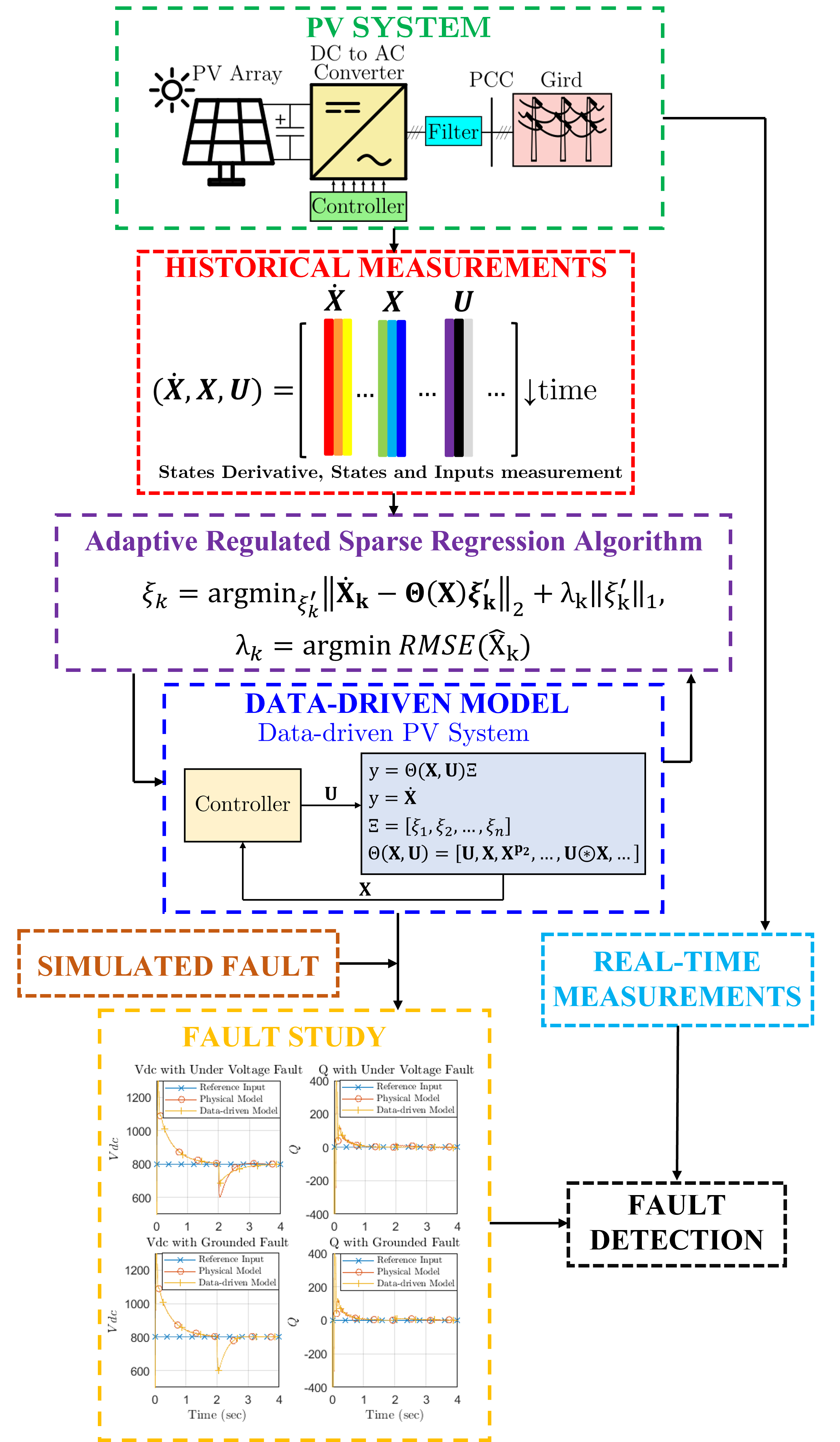}
    \caption{Fault study and fault detection of single-stage PV system.}
    \label{fsfd}
\end{figure}
\subsection{Fault Analysis Capability of ARSR Approach }
This case study introduces data-driven fault analysis capability of the ARSR approach in single-stage PV Systems. To ensure a fair and consistent comparison, all the reference inputs are set to be constants throughout this section.
Conducting fault tests on a real system entails significant risks, including the potential for component damage and increased costs. Therefore, conducting a simulated fault test using a data-driven model generated by the ARSR approach can help avoid damage to components and minimize costs. As depicted in Fig. \ref{fsfd}, the outcomes of the fault study find their ultimate application in fault detection by contrasting the response of a data-driven model with real-time measurements.
\begin{figure}[htb!]
    \centering
    \includegraphics[scale = 0.6]{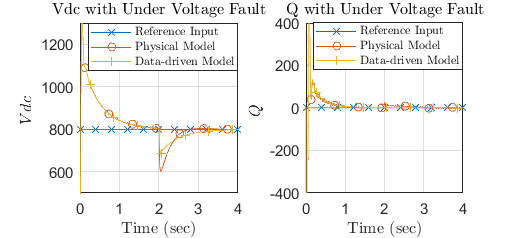}
    \caption{Reference tracking of physical and data-driven systems with under voltage fault.}
    \label{fl}
\end{figure}
Fig. \ref{fl} depicts the response of both physical  and data-driven models to an undervoltage fault condition. In this scenario, the grid voltage in the d-axis ($v_{gd}$) experiences a change from 800V to 500V at 3 seconds. The RMSE between the outputs of the physical model and the data-driven model for $v_{dc}$ and $Q$ are 1.0695 and 3.1428, respectively. Fig. \ref{f0} illustrates the comparison between a physical model and a data-driven model for a three-phase to ground fault. In this case, the grid voltage in the d-axis ($v_{gd}$) experiences a change from 800V to 0V at 3 seconds. The RMSE between the outputs of the physical model and the data-driven model for $v_{dc}$ and $Q$ are 1.0254 and 3.5588, respectively.
\begin{figure}[htb!]
    \centering
    \includegraphics[width = 0.45\textwidth]{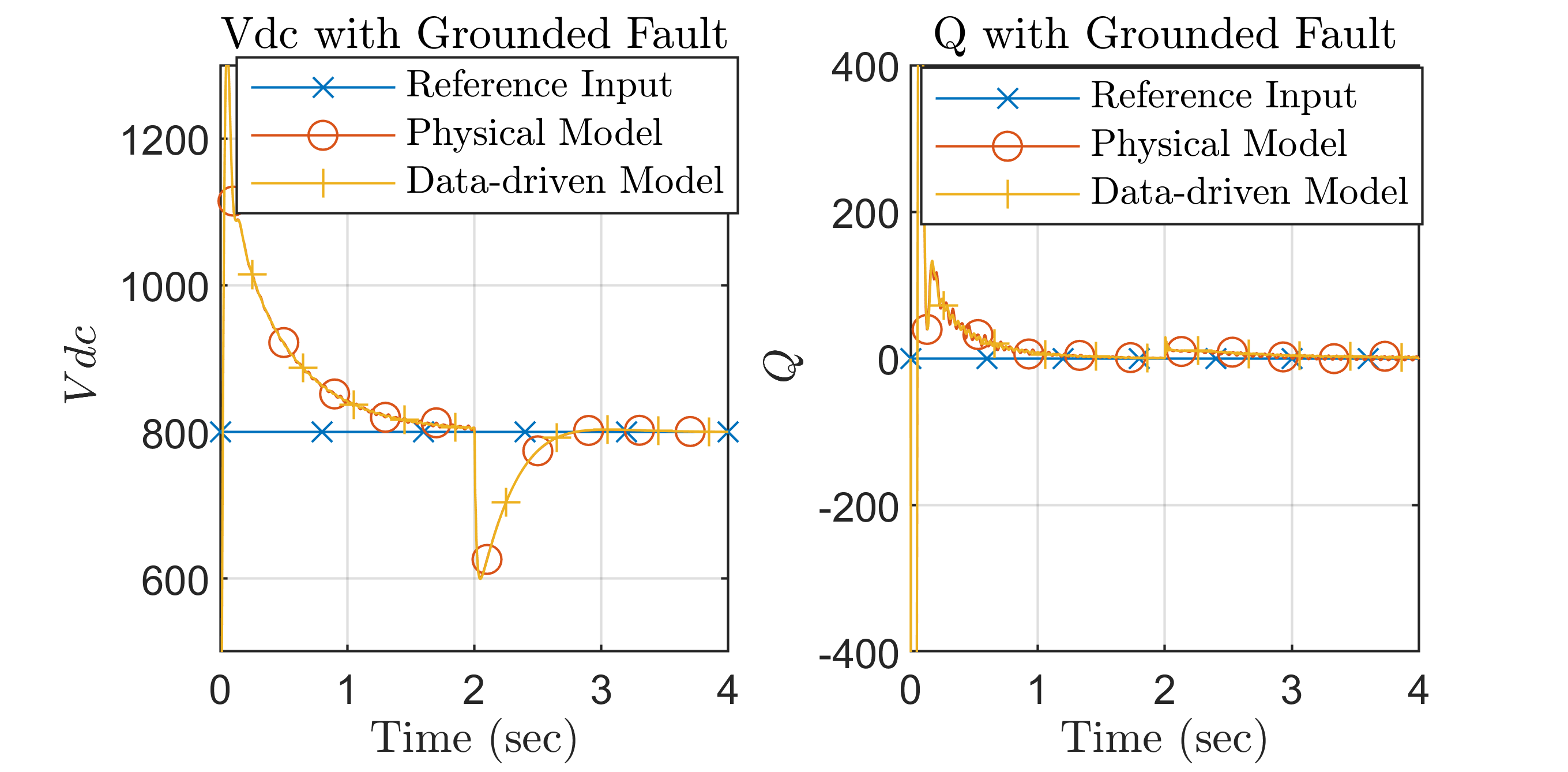}
    \caption{Reference tracking of physical and data-driven systems with three-phase grounded fault.}
    \label{f0}
\end{figure}
The results obtained from the fault tests in this section demonstrate that a data-driven model can provide insights into how a physical system reacts in a fault scenario. While it may not capture all the intricacies of the physical system, the data-driven model still offers valuable information about the system's behavior during faults.
\begin{figure}[htp!]
    \centering
    \includegraphics[width = 0.45\textwidth ]{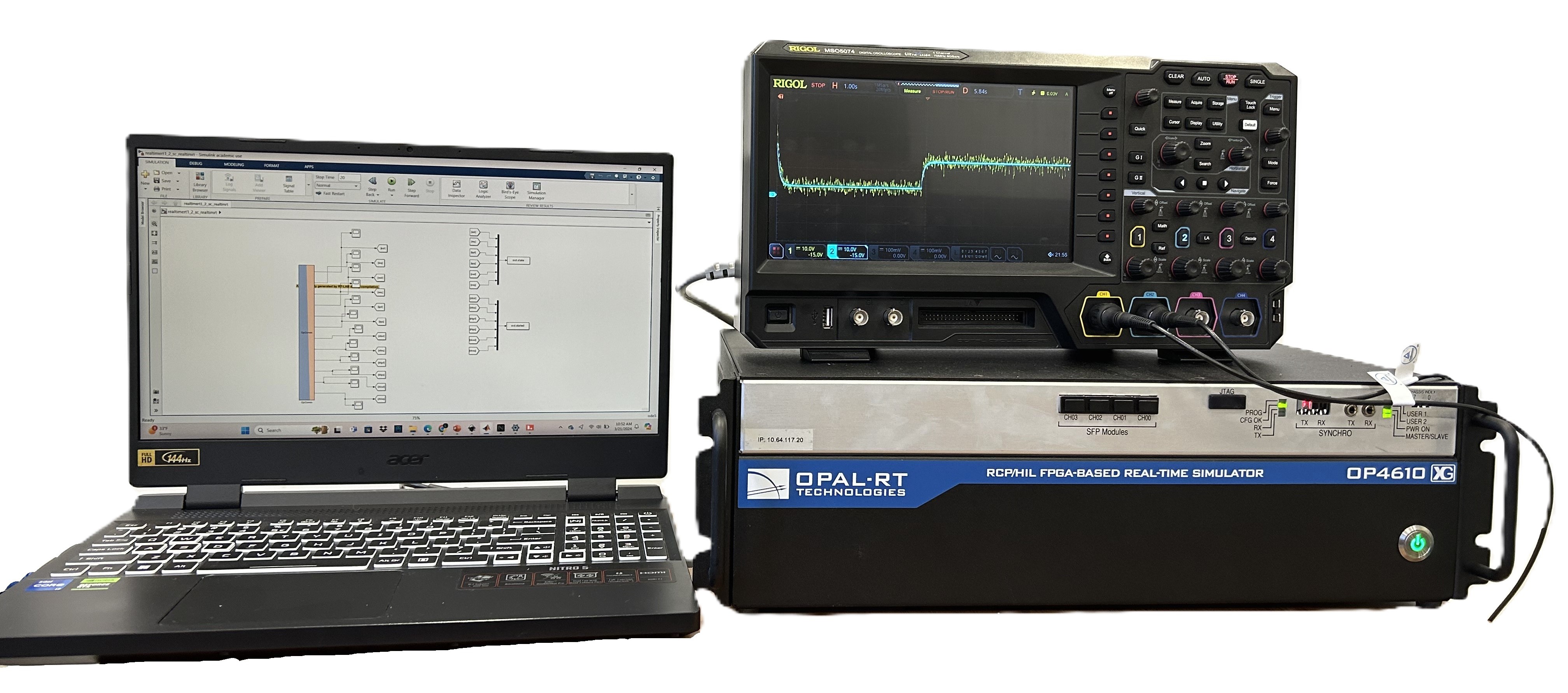}
    \caption{Real-time verification setup.  }
    \label{sco5}
\end{figure}
\begin{figure*}[htp!]
    \centering
    \includegraphics[width = 0.9\textwidth ]{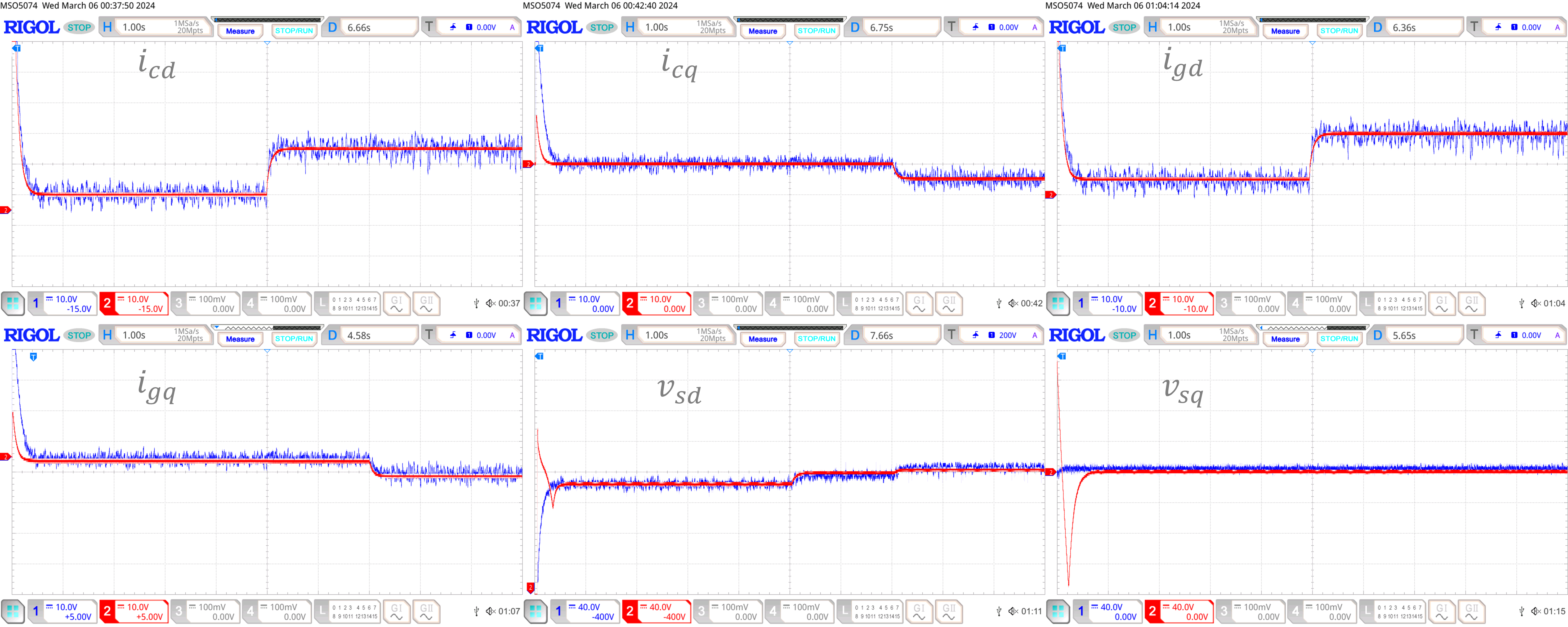}
    \caption{Real-time verification of 6 states in single-stage PV system.  }
    \label{sco}
\end{figure*}
\subsection{Real-time Simulation Verification}
This case study focuses on real-time simulation of the physical and data-driven models to verify the feasibility and performance of the proposed ARSR approach in real-world settings.
The validation process is conducted using the OPAL-RT OP4610XG real-time simulator. It involves comparing the single-stage PV system's physical model with its corresponding open-loop data-driven model, as described in Section II. Both the physical model and the data-driven model were constructed in MATLAB Simulink, were transformed into real-time models using RT-LAB software, and were run in the OP4610XG via an Ethernet connection. Subsequently, the simulation results are observed through the analog outputs of the OP4610XG using a digital oscilloscope, as shown in Fig.\ref{sco5}.
As depicted in Fig. \ref{sco}, screenshots captured from the oscilloscope display the measured and compared 6 states of the two models. The red curves represent the states of the data-driven model, closely matching the blue curves that represent the states of the physical model. This observation confirms the ARSR method's capability and feasibility in accurately replicating real-world physical systems.

\section{Conclusion}
This paper introduces ARSR algorithm and its application in data-driven modeling for single-stage and two-stage PV systems. Using measurements, sparse identification of nonlinear dynamics is employed to identify the dynamic models of PV systems. Open-loop and closed-loop data-driven models are developed for data-driven control of PV systems and the results are compared with physical models. Results confirm a close agreement with optimized mean squared error of prediction for the data-driven models. The proposed adaptive hyperparameter tuning approach increases the efficiency and accuracy of conventional sparse regression technique by adaptively identifying the hyperparameters that optimizing the error of prediction. Application of the proposed ARSR algorithm in fault study highlights the data-driven model's ability to simulate grid faults accurately. Finally, validations through real-time simulation confirm the applicability of the proposed approach for removing the dependency of existing PV controllers to known physical models. Future research will aim to create precise data-driven models in noisy environments, with a focus on exploring state estimation using the sparse regression approach in PV systems. 

\bibliographystyle{IEEEtran}
\bibliography{IEEEabrv,dp}

\begin{thebibliography}{10}
\providecommand{\url}[1]{#1}
\csname url@samestyle\endcsname
\providecommand{\newblock}{\relax}
\providecommand{\bibinfo}[2]{#2}
\providecommand{\BIBentrySTDinterwordspacing}{\spaceskip=0pt\relax}
\providecommand{\BIBentryALTinterwordstretchfactor}{4}
\providecommand{\BIBentryALTinterwordspacing}{\spaceskip=\fontdimen2\font plus
\BIBentryALTinterwordstretchfactor\fontdimen3\font minus
  \fontdimen4\font\relax}
\providecommand{\BIBforeignlanguage}[2]{{%
\expandafter\ifx\csname l@#1\endcsname\relax
\typeout{** WARNING: IEEEtran.bst: No hyphenation pattern has been}%
\typeout{** loaded for the language `#1'. Using the pattern for}%
\typeout{** the default language instead.}%
\else
\language=\csname l@#1\endcsname
\fi
#2}}
\providecommand{\BIBdecl}{\relax}
\BIBdecl

\bibitem{liu2020method}
M.~Liu, L.~Tan, and S.~Cao, ``Method of dynamic mode decomposition and
  reconstruction with application to a three-stage multiphase pump,''
  \emph{Energy}, vol. 208, p. 118343, 2020.

\bibitem{lu2020prediction}
H.~Lu and D.~M. Tartakovsky, ``Prediction accuracy of dynamic mode
  decomposition,'' \emph{SIAM Journal on Scientific Computing}, vol.~42, no.~3,
  pp. A1639--A1662, 2020.

\bibitem{al2021deep}
M.~Al-Gabalawy, ``Deep learning for koopman operator optimal control,''
  \emph{ISA transactions}, 2021.

\bibitem{mauroy2020koopman}
A.~Mauroy, I.~Mezi{\'c}, and Y.~Susuki, \emph{The Koopman Operator in Systems
  and Control: Concepts, Methodologies, and Applications}.\hskip 1em plus 0.5em
  minus 0.4em\relax Springer Nature, 2020, vol. 484.

\bibitem{sindy}
S.~L. Brunton, J.~L. Proctor, and J.~N. Kutz, ``Discovering governing equations
  from data by sparse identification of nonlinear dynamical systems,''
  \emph{Proceedings of the National Academy of Sciences}, vol. 113, no.~15, pp.
  3932--3937, 2016.

\bibitem{sindy_control}
U.~Fasel, E.~Kaiser, J.~N. Kutz, B.~W. Brunton, and S.~L. Brunton, ``Sindy with
  control: A tutorial,'' \emph{arXiv preprint arXiv:2108.13404}, 2021.

\bibitem{dmdp1}
G.~Kandaperumal, K.~P. Schneider, and A.~K. Srivastava, ``A data-driven
  algorithm for enabling delay tolerance in resilient microgrid controls using
  dynamic mode decomposition,'' \emph{IEEE Transactions on Smart Grid},
  vol.~13, no.~4, pp. 2500--2510, 2022.

\bibitem{koopp}
M.~Netto and L.~Mili, ``A robust data-driven koopman kalman filter for power
  systems dynamic state estimation,'' \emph{IEEE Transactions on Power
  Systems}, vol.~33, no.~6, pp. 7228--7237, 2018.

\bibitem{ddp1}
K.~A. Severson, P.~M. Attia, N.~Jin, N.~Perkins, B.~Jiang, Z.~Yang, M.~H. Chen,
  M.~Aykol, P.~K. Herring, D.~Fraggedakis \emph{et~al.}, ``Data-driven
  prediction of battery cycle life before capacity degradation,'' \emph{Nature
  Energy}, vol.~4, no.~5, pp. 383--391, 2019.

\bibitem{li2022neural}
Y.~Li, Y.~Liao, X.~Wang, L.~Nordstr{\"o}m, P.~Mittal, M.~Chen, and H.~V. Poor,
  ``Neural network models and transfer learning for impedance modeling of
  grid-tied inverters,'' in \emph{2022 IEEE 13th International Symposium on
  Power Electronics for Distributed Generation Systems (PEDG)}.\hskip 1em plus
  0.5em minus 0.4em\relax IEEE, 2022, pp. 1--6.

\bibitem{syndy}
S.~L. Brunton, J.~L. Proctor, and J.~N. Kutz, ``Discovering governing equations
  from data by sparse identification of nonlinear dynamical systems,''
  \emph{Proceedings of the national academy of sciences}, vol. 113, no.~15, pp.
  3932--3937, 2016.

\bibitem{syndy1}
------, ``Sparse identification of nonlinear dynamics with control (sindyc),''
  \emph{IFAC-PapersOnLine}, vol.~49, no.~18, pp. 710--715, 2016.

\bibitem{10372091}
A.~Hosseinipour and J.~Khazaei, ``Sparse identification for data-driven
  dynamics and impedance modeling of power converters in dc microgrids,''
  \emph{IEEE Journal of Emerging and Selected Topics in Industrial
  Electronics}, pp. 1--13, 2023.

\bibitem{ddsparse}
J.~Khazaei and A.~Hosseinipour, ``Data-driven feedback linearization control of
  distributed energy resources using sparse regression,'' \emph{IEEE
  Transactions on Smart Grid}, pp. 1--1, 2023.

\bibitem{msindy}
A.~Nandakumar, Y.~Li, H.~Zheng, J.~Zhao, D.~Zhao, Y.~Zhang, T.~Hong, and
  B.~Chen, ``Data-driven modeling of microgrid transient dynamics through
  modularized sparse identification,'' \emph{IEEE Transactions on Sustainable
  Energy}, pp. 1--14, 2023.

\bibitem{ddnaval}
\BIBentryALTinterwordspacing
J.~Khazaei and A.~Hosseinipour, \emph{Advances in Data-Driven Modeling and
  Control of Naval Power Systems}.\hskip 1em plus 0.5em minus 0.4em\relax John
  Wiley \& Sons, Ltd, 2022, ch.~21, pp. 453--473. [Online]. Available:
  \url{https://onlinelibrary.wiley.com/doi/abs/10.1002/9781119812357.ch21}
\BIBentrySTDinterwordspacing

\bibitem{gpe}
A.~Hamid, D.~Rafiq, S.~A. Nahvi, and M.~A. Bazaz, ``Power grid parameter
  estimation using sparse identification of nonlinear dynamics,'' in \emph{2022
  International Conference on Intelligent Controller and Computing for Smart
  Power (ICICCSP)}, 2022, pp. 1--6.

\bibitem{single}
V.~N. Lal and S.~N. Singh, ``Control and performance analysis of a single-stage
  utility-scale grid-connected pv system,'' \emph{IEEE Systems Journal},
  vol.~11, no.~3, pp. 1601--1611, 2017.

\bibitem{two}
J.~Khazaei, Z.~Tu, and W.~Liu, ``Small-signal modeling and analysis of virtual
  inertia-based pv systems,'' \emph{IEEE Transactions on Energy Conversion},
  vol.~35, no.~2, pp. 1129--1138, 2020.

\bibitem{iravanibook}
A.~Yazdani and R.~Iravani, \emph{Voltage-sourced converters in power systems:
  modeling, control, and applications}.\hskip 1em plus 0.5em minus 0.4em\relax
  John Wiley \& Sons, 2010.

\end{thebibliography}

\end{document}